\documentstyle[twocolumn,aps,floats,amssymb,epsfig]{revtex}
\begin{document}

\def\d{{\rm d}}
\def\e{{\rm e}}
\def\i{{\rm i}}
\def\O{{\rm O}}
\def\half{\mbox{$\frac12$}}
\def\eref#1{(\protect\ref{#1})}
\def\etal{{\it{}et al.}}
\def\Li{\mathop{\rm Li}}
\def\av#1{\left\langle#1\right\rangle}
\def\set#1{\left\lbrace#1\right\rbrace}
\def\stirling#1#2{\Bigl\lbrace{#1\atop#2}\Bigr\rbrace}

\draft
\tolerance = 10000

\renewcommand{\topfraction}{0.9}
\renewcommand{\textfraction}{0.1}
\renewcommand{\floatpagefraction}{0.9}
\setlength{\tabcolsep}{4pt}

\twocolumn[\hsize\textwidth\columnwidth\hsize\csname @twocolumnfalse\endcsname

\title{Intentional Walks on Scale Free Small Worlds}
\author{ Amit R Puniyani$^{1}$,
Rajan M Lukose$^{2}$ and Bernardo A. Huberman$^{2}$}
\address{$^1$
Department of Physics, Stanford University, Stanford CA 94305}
\address{$^2$ HP Labs, 1501 Page Mill Road, CA 94304-1126}
\maketitle

\begin{abstract}

We present a novel algorithm that generates scale free small world
graphs such as those found in the World Wide Web,social and metabolic
networks. We use the generated graphs to study the dynamics of a
realistic search strategy on the graphs, and find that they can be
navigated in a very short number of steps.
\end{abstract}
]
\newpage

Small world and scale free graphs, which are at the heart of
systems as diverse as the World Wide Web \cite{ladamicsw}, the
call logs of telephone networks \cite{chung}, social and
professional acquaintances
\cite{newman_social,wattssw,strogatz_review,newman_review}, power
grids \cite{wattssw} and metabolic networks \cite{barabasi,fell},
have attracted a lot of attention from the physics community in
recent years (see the reviews
\cite{barabasi_review,newman_review}).

One reason for the interest has been the number of dynamical
processes such as percolation
\cite{callaway,moore,cascade,huberman_percolation}, epidemic spreading \cite{pastor,moore,wattssw}, random walks
\cite{pandit} and  message-passing \cite{kleinberg} which are of
fundamental importance to statistical physics and have numerous
applications in areas such as robustness of the Internet and the
power grid, the spread of epidemics in societies, the spread of
computer viruses in computer networks, routing in large computer
networks \cite{upfal} and in measuring the efficiency of online
algorithms which utilize (www) network topology \cite{stoch_web}.

We present a novel algorithm that generates scale free small world
graphs such as those found in the World Wide Web,social networks and metabolic
networks. We use the generated graphs to study the dynamics of a
realistic search strategy on the graphs, and find that they can be
navigated in a very short number of steps.

Watts and Strogatz \cite{wattssw} first developed a procedure for
generating graphs which have both short path lengths and
clustering. This was an improvement over traditional Erdos-Renyi
random graphs \cite{erdos}. The Watts-Strogatz procedure however,
lacks an important property exhibited by social and other
networks, i.e. their approximate power-law distribution in the
number of a node's links. This distribution amounts to stating
that a few nodes or people or sites in the web have very many
links whereas most have a few. Whereas there are some small world
graphs that are not power-law like (e.g. the electric power grid),
many are scale free, such as the call graph of large-scale
telephone use, the Web, the Internet backbone, and metabolic
networks.

Recently, Barabasi et al \cite{barabasi2}, described a procedure
for producing random graphs with a power-law distribution while
failing to produce graphs that also have the clustering property
of small worlds. While this work generates networks analogous to
the power grid, it fails at generating the clustering property
known to exist in the link structure of the World Wide Web \cite{ladamicsw},
metabolic networks \cite{fell} or social networks \cite{newman_social}.

The issue of navigation also received a partial answer in a paper
by Kleinberg \cite{kleinberg}, who used a 2-D lattice substrate
and a regular distribution of links. Motivated by real experiments
with social networks, Kleinberg was concerned with how, given the
fact that short paths existed, one could find them without
complete global information.  The treatment given in
\cite{kleinberg} had an elegant result, but the underlying graph
model did not reflect all of the important features real world
problems. An important shortcoming is its particular assumption of
an inverse square correlation that implies that a majority of ones
contacts lie in close geographical proximity. What happens if a
large fraction of people know as many people outside of their city
or state as inside? Would it become impossible to pass messages
efficiently? What happens if the graph representing the social
network, cannot be embedded on a two-dimensional lattice?  Is it
possible to devise an optimal strategy to navigate these networks?
\vskip 0.2 cm

In this paper we solve all these shortcomings by presenting a
general procedure for constructing small-world graphs with
power-law distribution in their link structure, and a robust
near-optimal strategy for navigating such small-world graphs.

\vskip 0.2 cm

We start by constructing small-world graphs with power-law
distributions and large clustering coefficients. The clustering
coefficient is the probability that two links connected to a
common node have a link. In real social networks, clustering
coefficients as high as 0.2 are not surprising \cite{strogatz_review,wattssw}.

\vskip 0.2 cm

To do so we first assign each node with a number of links given by
a power law distribution i.e. We compute from a truncated power
law function the number of nodes say $n$ having a given number of
links $l$. The maximum number of links a given node has is a
certain large number $l$.  We then choose $n$ nodes at random from
the ring and label them with $l$. Then we connect nodes assigned a
given degree with their nearest unsaturated neighbors on the ring.
(An unsaturated neighbor is a neighbor who has not been linked to
as many links as assigned in the first step).

\vskip 0.2 cm

At this point, if we randomly rewire the graph in the
Watts-Strogatz way, the power-law topology would be destroyed. To
obtain a small-world graph with a power-law distribution from a
graph that we just formed, it is necessary to keep the number of
nodes with a given number of links a constant. We achieve this by
adding and deleting links in a way that the total number of links
at each node is conserved. We choose a node $A$ at random, delete
a link say with $D$ and then immediately form a link between this
node and a new node $B$ chosen at random to conserve links at $A$.
Then we delete a link from $B$, say the one with node $C$. This
deletion will cause $C$ to have one link less and will conserve
the links at $B$. This is what the situation was at the first
deletion so putting $A=C$ we repeat the process. Note that except
for $D$ all the other nodes have the same number of links as
before and yet there are shortcuts in the system. \vskip 0.2 cm

The rewiring procedure and the graphs generated are illustrated in
Fig 1.

\vskip 0.5 cm

\epsfysize=20mm \epsfbox{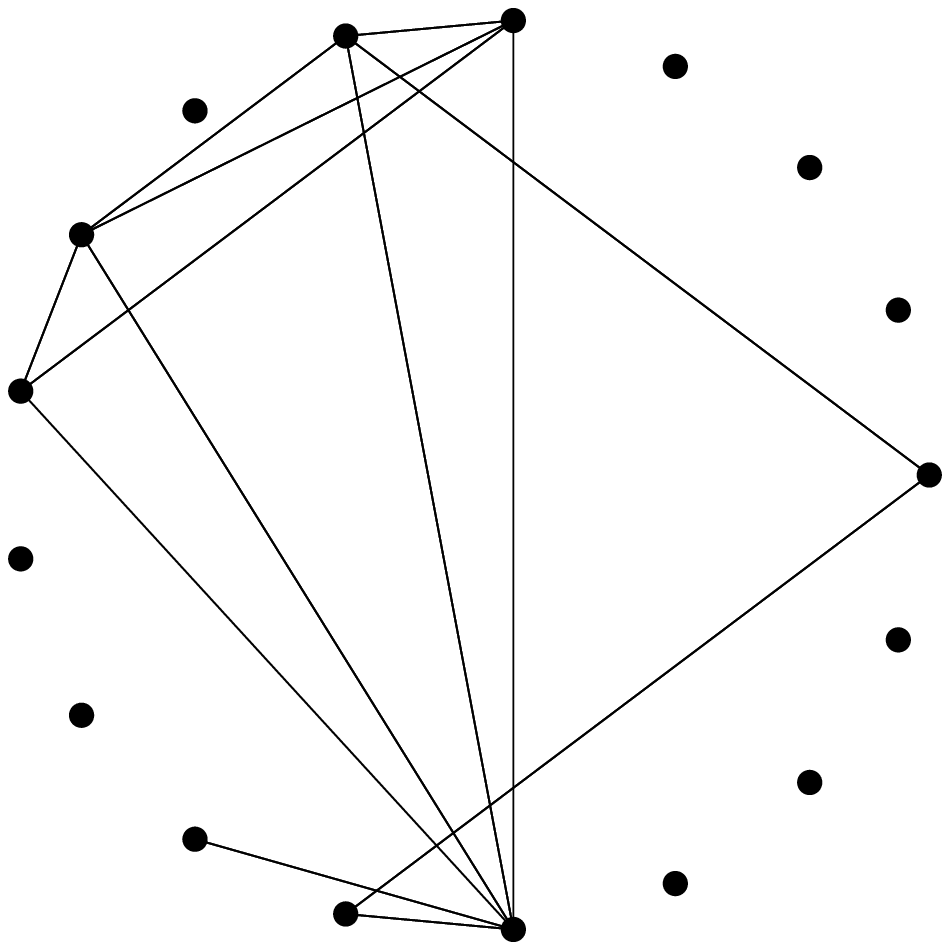}
\epsfysize=20mm \epsfbox{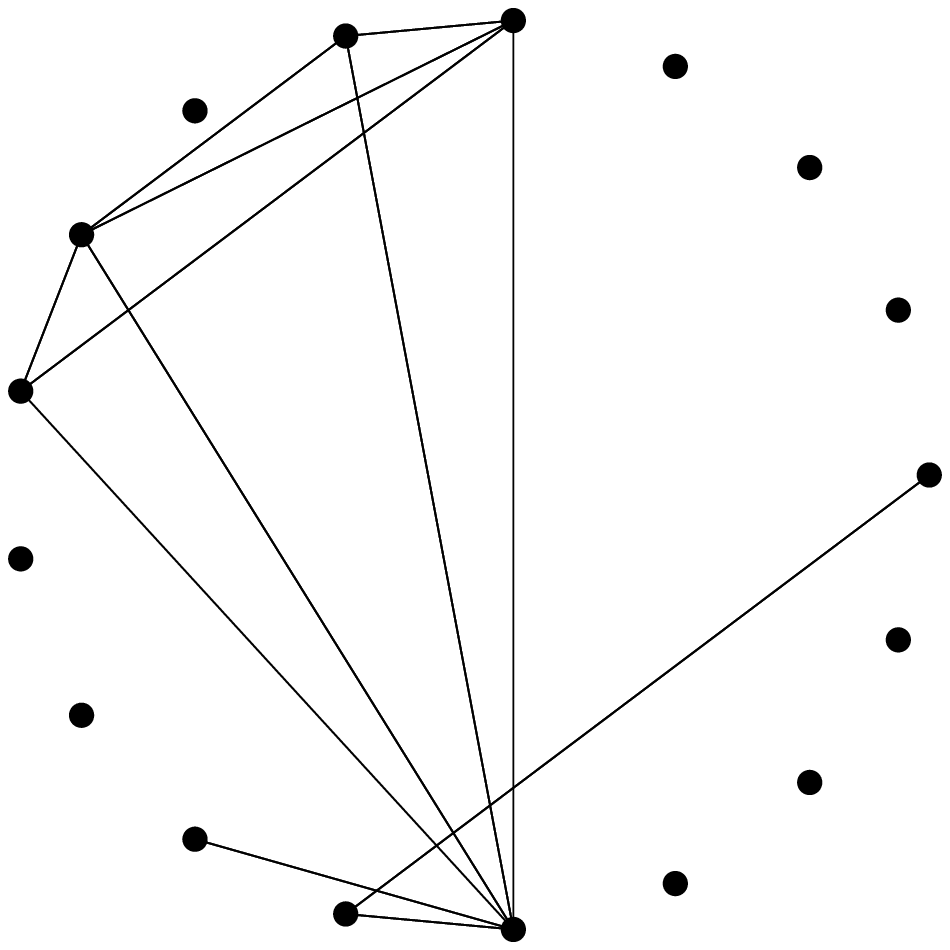}
\epsfysize=20mm \epsfbox{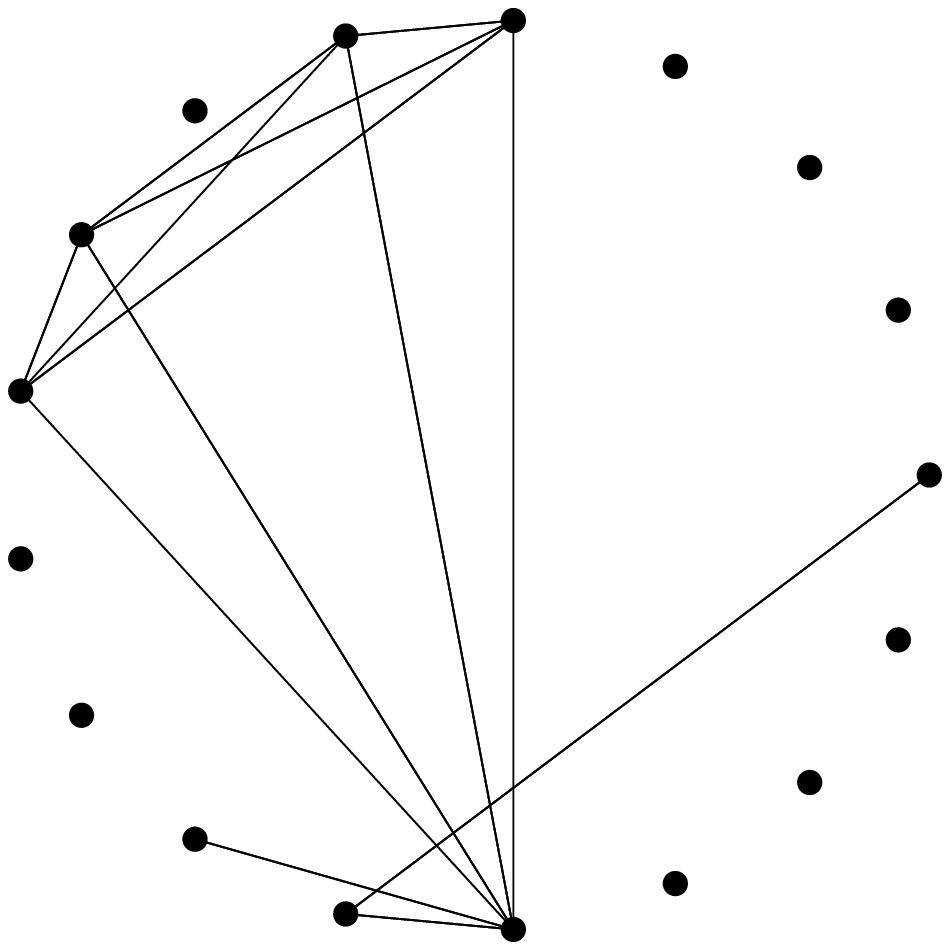}
\epsfysize=20mm \epsfbox{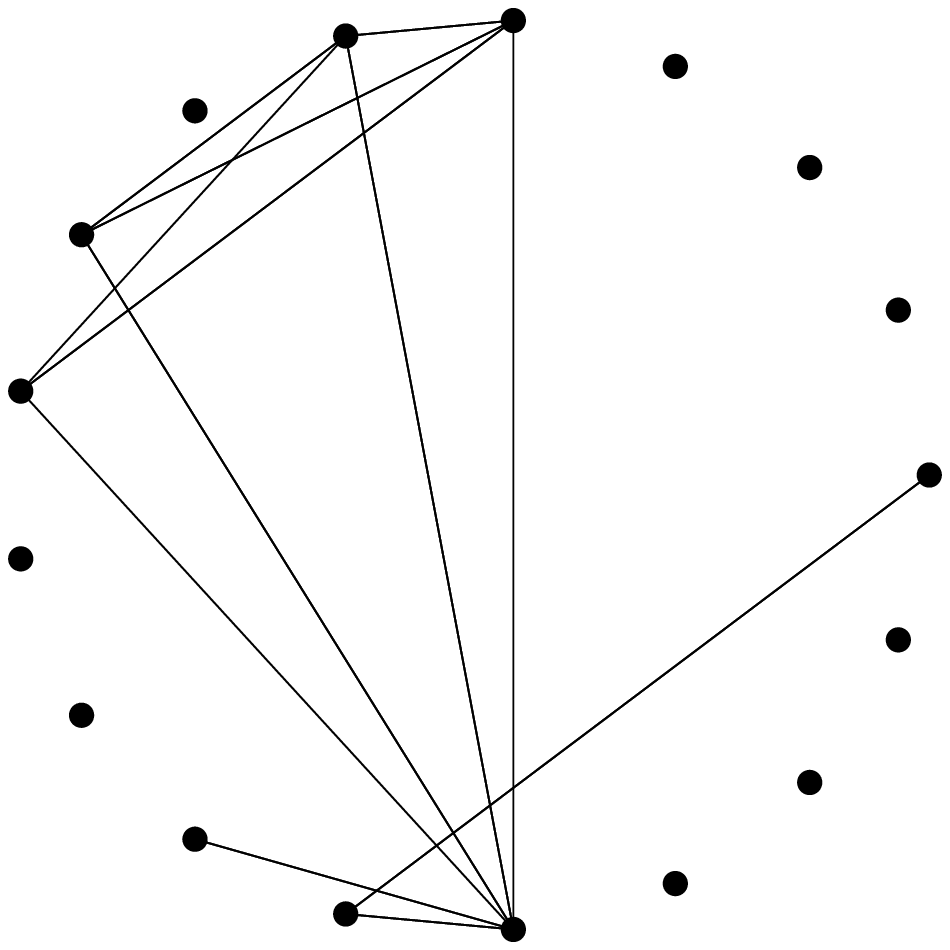}
\vskip 0.2 cm
\epsfysize=25 mm \epsfbox{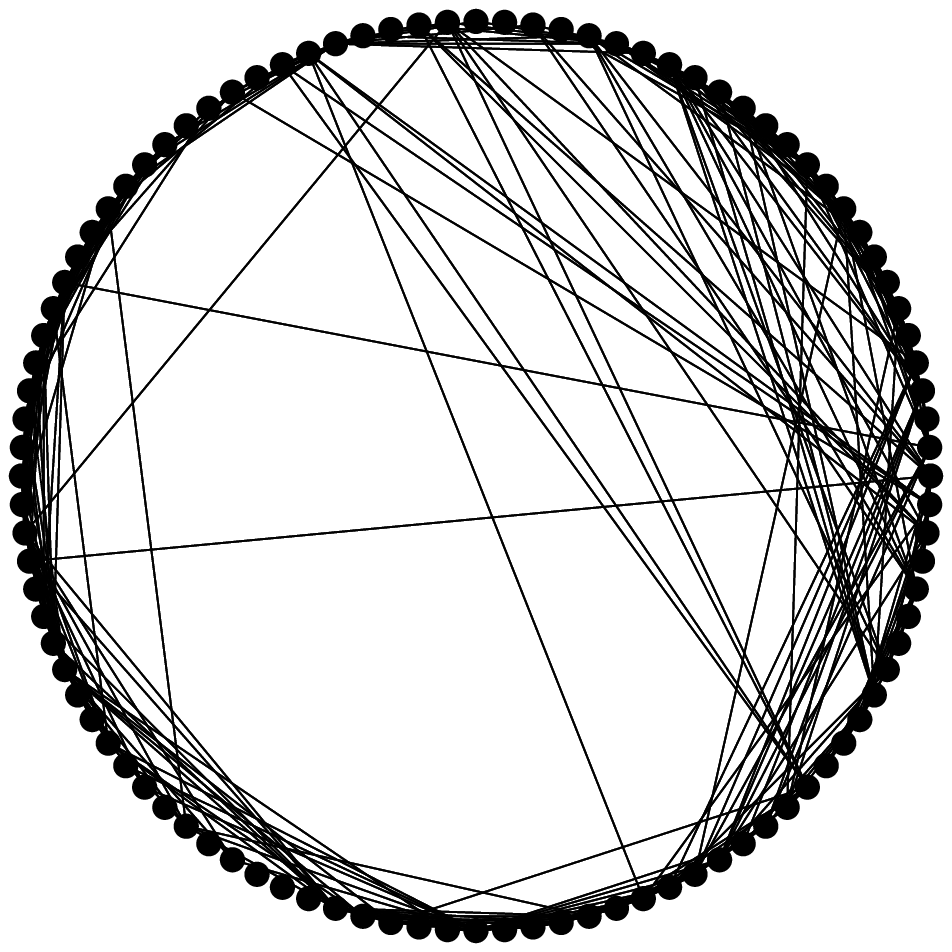} \hskip 0.4 cm \epsfysize=25 mm \epsfbox{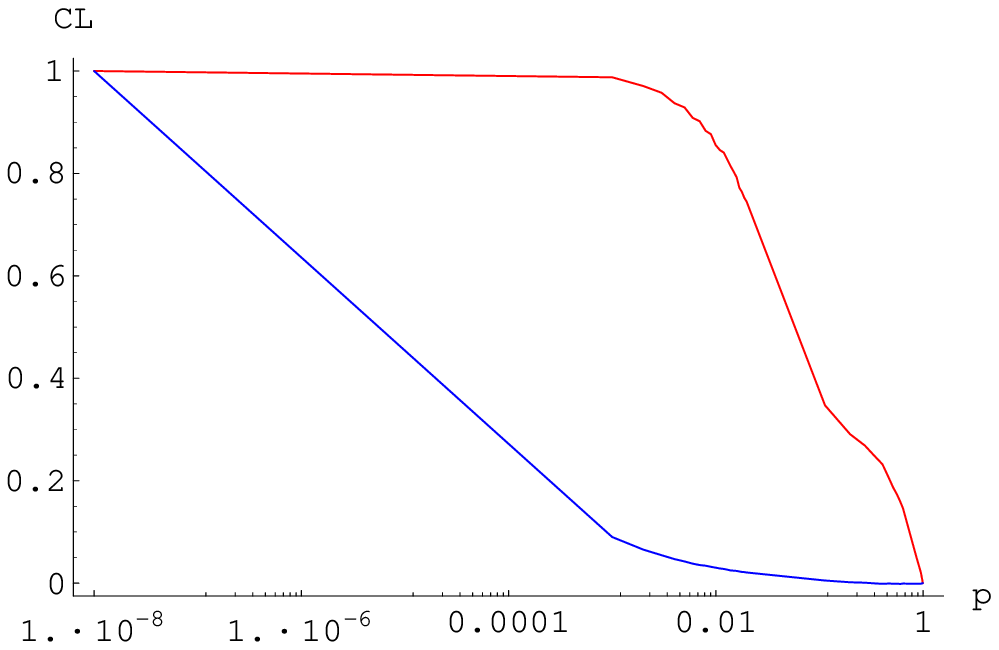}
\vskip 0.3 cm
{\it Figure 1: Illustration of rewiring procedure. A link is removed and a
new one added such that every node has the same number of links. Graph generated with 300 nodes and power 2 ie. the
number of nodes with l links goes as $\frac{1}{l^{2}}$ and Plot showing
clustering coefficient(red) and average path length(blue) as
a function of the randomisation. Note that the path length comes down
drastically while the clustering coefficient persists close to it's
value for zero randomisation }
\vskip 0.5 cm

By increasing the fraction of links rewired we get the required
short path length. If the fraction of links deleted and rewired is
$p$, then for very small $p$ the average path length $L(p)$comes
down by orders of magnitude and is close to that of the
corresponding random graph whereas the clustering coefficient
$C(p)$ is still much larger than the corresponding random graph
value. The plots of $L(p)$ and $C(p)$ are shown in Fig 1. \vskip
0.5 cm

This procedure can generate graphs with clustering coefficients
between 0 and 0.1 and arbitrary power-law exponents and path
lengths close to logarithmic. The procedure can also be modified
for directed graphs. The flexibility makes it possible to study
various dynamical properties of real systems by doing simple
simulations on these models.

Next we study navigation strategies which are of relevance to the
study of social networks, \cite{milgram,kleinberg} and also to the
computer networks \cite{upfal}. In both cases it is necessary to
generate graphs with power law link distribution with exponent around 2
and with clustering of the order of 0.1 as is observed in social
\cite{chung} and computer networks \cite{diameter}.

We consider a navigation strategy that can be described as an
"intentional walk", the walker {\it intending} to reach a target
node. All nodes are labelled by a distinct co-ordinate and the
co-ordinates of the target are known to the walker. The walker
must use only local information such as the co-ordinates of the
node it is at and it's neighbors to take the next step. The walk
ends when the walker reaches the target.

In the social context or in the computer networks context, this is
equivalent to passing a message or packet through the network towards a
specified destination using as little information about the global
connectivity map as possible.

This kind of walk does not amount to a pure random walk, since the
nodes forwarding the message, while not having detailed knowledge
of the network in which they are embedded, do have a sense of
direction when forwarding these messages.

Milgram first studied it empirically \cite{milgram} in social
Networks. Milgram found that knowledge of geographical location of
the target helped the messages find very short paths to the
target. This has been dubbed in the popular media as the 'Small
World Phenomenon' or 'Six Degrees of Separation'. Kleinberg
\cite{kleinberg} studied this navigation problem analytically
using a limited and particular graph model, in which the link
distribution is unrealistic.

In contrast to Kleinberg, we study this problem on scale-free
small worlds, generated by the above algorithm. We present a novel
and realistic strategy for navigating these small-worlds in a near
optimal fashion, i.e. that given any source and target node the
number of steps needed to reach an arbitrary target is close to
the shortest possible.

The strategy uses partial
knowledge that each person/node has about the positions(along the
ring) of its next neighbors. For example, if  person $A$ in
California needs to deliver a letter (the intentional walker) to
someone who lives in a school campus in Boston but has no contacts
in Boston, she would send it to an acquaintance whom she knows
visits Boston on a regular basis. This is to be contrasted with
the strategy chosen in Kleinberg's work which resorts to physical
propinquity, i.e. send to someone who is close to Boston.

As we show, this kind of knowledge implemented as a strategy leads
to a drastic speed up in the journey of an intentional walker to
the target, to the point of making it near-optimal, i.e. the
intentional walker almost always takes a number of steps equal to
or a small fraction of more than the shortest path between source
and target nodes in the network.

We label the nodes by their position along the ring and the
position of the target is known. However, the exact path or paths
which lead to the target are unknown.

The results for this algorithm for a graph with 10
Power Law of exponent 2.1 and Clustering coefficient C=0.08 are
shown in Figure 2 (a). The results show that the scaling of the number of
steps (search cost) with size is logartihmic.

\vskip 0.5 cm
\epsfysize=25 mm \epsfbox{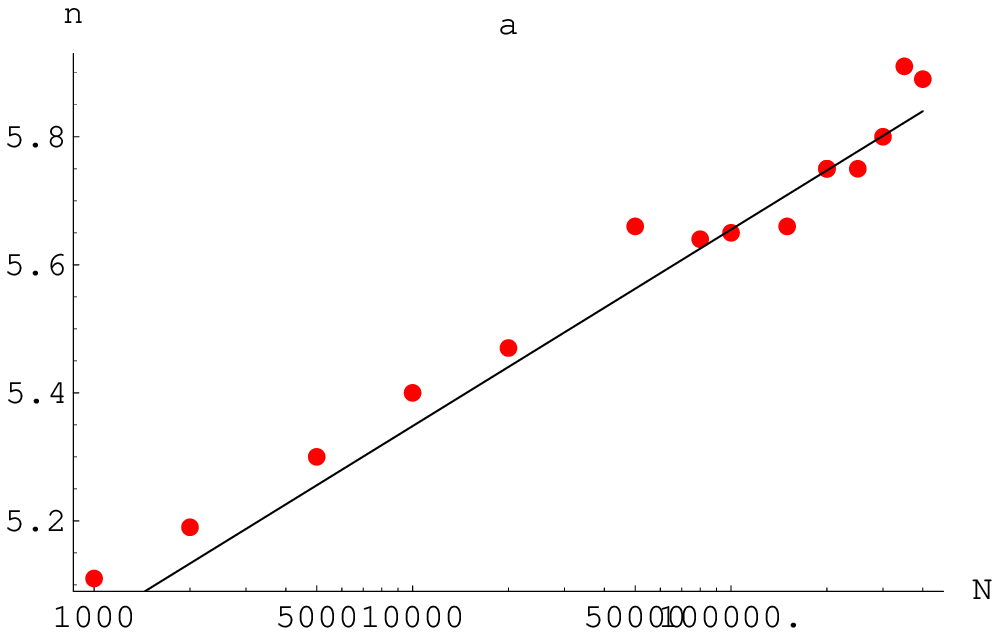}
\hskip 0.2cm
\epsfysize=25 mm \epsfbox{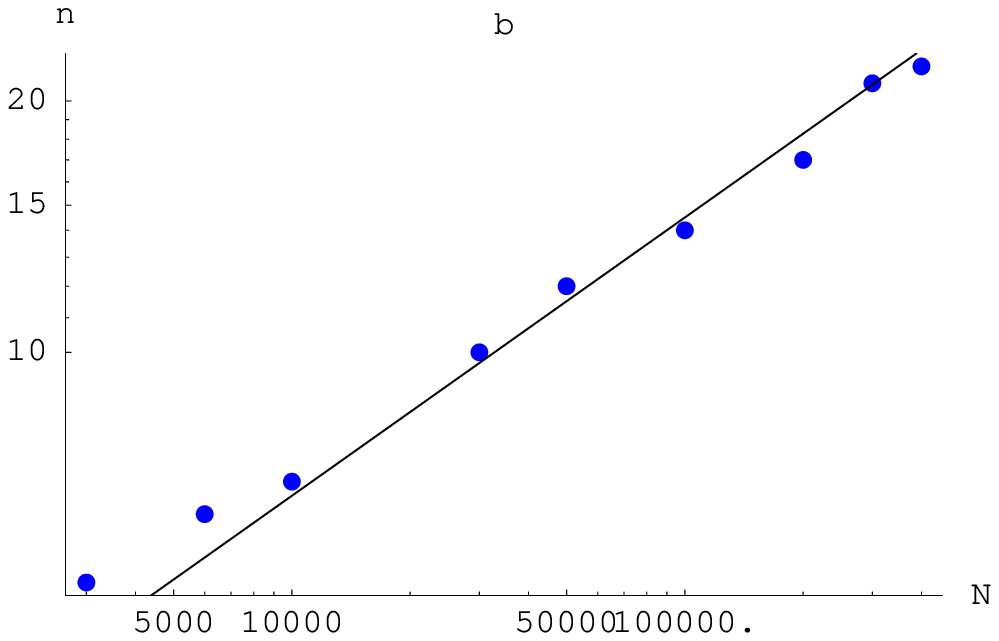}
\vskip 0.3 cm
\epsfysize=25 mm \epsfbox{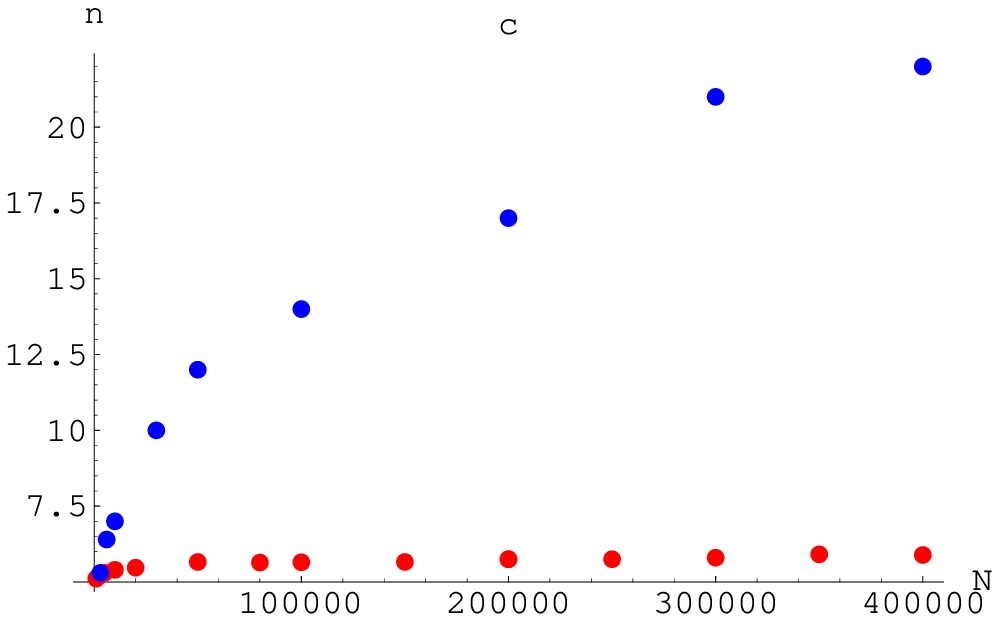} \hskip 0.2 cm \epsfysize=25
mm \epsfbox{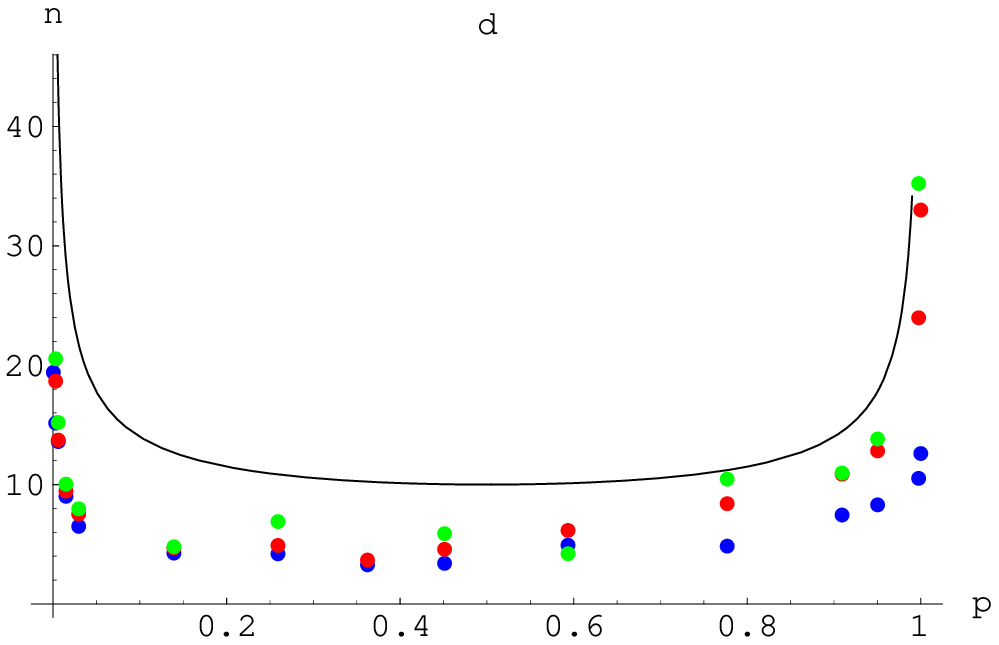}
\vskip 0.5cm
{\it Figure 2 (a): showing the average number of steps n(y-axis)
it takes to reach the target for a power law graph (exponent 2.1 with
clustering coefficient C=0.08)  of size N (x-axis)
 and the fit to a logarithmic function n=2/15*Log(N)+4.12  on a Log-linear
plot.
 2 (b): showing the same for Watts Strogatz graph. Here
$n=\frac{1}{3.2}*N^{1/3}$ , 2(c) showing the comparison between the SWPL (red)
and the WS-SW (blue) and 2(d) showing the variation of the convergence time in a PL-SW as a function
of p the fraction of random links for N=1000 (blue), N=10000(red) and N=100000(green) and the upper
bound function (equation (4)) (with $\alpha=9$) in black }
\vskip 0.5 cm

We performed intentional walking experiments on Watts Strogatz
type of Small Worlds (WS-SW) . The results of
the WS-SW message passing experiments are shown in Figure 2 (b). It is
linear on a Log-Log plot indicating that the scaling is polynomial.

The search cost for the intentional walk on the
WS-SW (Watts Strogatz Small World)scales polynomially ($N^{1/3}$ meaning that for a
graph of size 100 million (which is the order of magnitude of the population of the US) the number of steps it takes is of the order of
100. For the PL-SW (Power Law Small World) , the search is much more
efficient, the cost (the number of steps needed to find the target)
scaling  logarithmically with the size.
When
extrapolated the function giving the number of steps for the PL-SW would
equal 6.5 for a graph size 100 million. This is very close to what
Milgram observed in his experiment on the American population \cite{milgram}.

The search costs for a PL-SW and a WS-SW for approximately the same
average degree are plotted together in Figure 2(c). Note that the difference in the search cost is not because there are no
short paths in the case of the WS-SW. In fact the WS-SW has path lengths
which scale logarithmically with the size \cite{newman} just as the
PL-SW. It is just that algorithms utilizing local information are unable
to find them.

All this indicates that a realistic link-distribution plays a
crucial role in the effectiveness of the strategy and that it
might be a crucial ingredient in explaining the small numbers seen
in message passing experiments on social networks such as
Milgram's \cite{milgram}

The basic principle behind the discovery of short paths is that in
a power-law graph the expected degree of a node following an edge
is much larger than the average degree \cite{newman}. This means
that each node is connected to some high degree nodes. Thus there
are many second neighbors.  Most of the second neighbors would be
local in a small-world but a finite fraction would be randomly
distributed throughout the network. Since there would be so many
second neighbors, with high probability one of those randomly
placed ones would be located close to the target.

The intentional walker can thus get close to the target in a small number
of steps. On getting close it would find the local links more efficient in
taking it to it's target. It is thus a combination of the properties of scale free ness and
the existence of local links (clustering) and of course the fact that a
finite fraction of the links are randomly.

If the expected degree of a neighbor is E and the fraction of
random links is p, then it would have approximately $E^{2}$ second
neighbors out of which p would be long range or random links so
the average time it would take to reach the target if it were to
never use local links is:
$$n_{1}=\frac{N}{E^{2}p}+\beta\eqno{(1)}$$

where $\beta$ is the average minimum number of steps required to reach the
target. Note that it is impossible for the walker to reach  in less than
3 steps according to this procedure, independent of the size. This is
because the the average shortest path of a graph has a finite lower bound
according to Newman's formula for graph diameter \cite{newman}. In other
words the plot of diameter with size has a finite y-intercept which is
slightly more than  3 for most powers.

According to our algorithm, the intentional walker would go to the
closest node to the target among it's second and first neighbors
every 2 steps. It becomes advantageous to use local as opposed to
random links once it has come close to the target say a distance
$l$. The point at which this happens is such that the total number
of steps is minimal with respect to $l$. Note that jumping along
local links is like jumping along a ring in the worst case and the
time it takes to reach the target by this kind of jumping is
linear in the size of the total distance. The maximum number of
steps it takes if the walker starts using local links after
reaching within a distance $l$ of the target will be following (1):

$$n=\frac{N}{E^{2}l p}+ \frac{\alpha l}{\left(1-p \right)}+ \beta \eqno{(2)}$$
where $\alpha$ and $\beta$ are constants. For a power law graph of
power approximately 2 as observed in most networks of interest
\cite{chung,barabasi,diameter} we have the expression for $E$ in
terms of $N$,

$$E=\frac{N^{1/2}}{Log (N)} \eqno{(3)}$$

Now, substituting (3) and minimizing the resulting expression
with respect to $l$ we get an upper bound for the average number
of steps:

$$n=\sqrt{\frac{\alpha}{p\left(1-p\right)}} Log(N)+\beta \eqno{(4)}$$

We plot this with respect to $p$, the fraction of random links (for
average $degree=13$ and $N=100000$,
$\beta=3.4$) and also plot the actual time for a series of
different $N$ in Figure 2 (d). Note that the most efficient search is
when there are both local and random long distance links.

To conclude we have developed a method that can generate real
world-like graphs which have short path lengths, clustering and
power law link distributions. This can be extended to directed
graphs and arbitrary link distributions. It is possible to use
this procedure to generate graphs in order to study scaling
properties of various algorithms which use the link distribution
of a network. We illustrated this with a study of the classic
message passing or intentional walk problem on scale free small
worlds (with power law of 2.1, C=0.1 like in social and computer
networks where this problem is important). We estimated the
efficiency of a novel intentional walk algorithm utilizing second
neighbor information using this procedure.  The ability to
construct realistic graph models easily will enable better
characterization of the structure and dynamics of important
large-scale network systems.

\bibliography{milgram3}

\begin{thebibliography}{10}

\bibitem{ladamicsw}
L.~Adamic.
\newblock The small world web.
\newblock {\em ECDL}, 1999.

\bibitem{chung}
W.~Aiello, F.~R.~K. Chung, and L.~Lu.
\newblock A random graph model for massive graphs.
\newblock {\em {ACM} Symposium on Theory of Computing}, pages 171--180, 2000.

\bibitem{barabasi_review}
R.~ALbert and A.-L. Barabasi.
\newblock Statistical mechanics of complex networks.
\newblock {\em cond-mat/0106096}.

\bibitem{callaway}
D.S.Callaway, M.~Newman, S.H.Strogatz, and D.~Watts.
\newblock Network robustness and fragility: Percolation on random graphs.
\newblock {\em Physical Review Letters}, 85:5468--5471, 2000.

\bibitem{barabasi2}
A.-L.~B. et~al.
\newblock Scale free charecter of random networks: the topology of the www.
\newblock {\em Physica A}, 281:69--77, 2000.

\bibitem{diameter}
H.~Jeong, R.~Albert, and A.~L. Barabasi.
\newblock Diameter of the world wide web.
\newblock {\em Nature}, 401:130, 1999.

\bibitem{barabasi}
H.~Jeong, B.~Tombor, R.~Albert, Z.~Oltvai, and A.-L.Barabasi.
\newblock The large-scale organization of metabolic networks.
\newblock {\em Nature}, 407:651--654, October 2000.

\bibitem{kleinberg}
J.~Kleinberg.
\newblock Navigation in a small world.
\newblock {\em Nature}, 406, 2000.

\bibitem{milgram}
S.~Milgram.
\newblock The small world problem.
\newblock {\em Psychology Today}, 1:61, 1967.

\bibitem{moore}
C.~Moore and M.~Newman.
\newblock Epidemics and percolation in small world networks.
\newblock {\em Phsical Review E}, 61:5678--5682, 2000.

\bibitem{newman_review}
M.~Newman.
\newblock Models of the small world.
\newblock {\em Journal of Statistical Physics}, 101:819--841, 2000.

\bibitem{newman_social}
M.~Newman.
\newblock The structure of scientific collaboration networks.
\newblock {\em Proc. Natl. Acad. Sci}, 98:404--409, 2001.

\bibitem{newman}
M.~E.~J. Newman, D.~J. Watts, and S.~H. Strogatz.
\newblock Random graphs with arbitrary degree distribution.
\newblock {\em condmat}, 0104209, 2001.

\bibitem{pandit}
S.~Pandit and R.~E. Amritkar.
\newblock Random spread on the family of small-world networks.
\newblock {\em cond-mat}, 0004163, 2000.

\bibitem{pastor}
R.~Pastor-Satarros and A.~Vespignani.
\newblock Epidemic spreading in scale free networks.
\newblock {\em Physical Review Letters}, 86:3200, 2001.

\bibitem{upfal}
D.~Peleg and E.~Upfal.
\newblock A trade-off between size and efficiency for routing tables.
\newblock {\em Communications of the ACM}, 30:3, 1997.

\bibitem{erdos}
P.Erdos and A.Renyi.
\newblock On the evolution of random graphs.
\newblock {\em Pupl. Math. Inst. Hungar. Acad. Sci.}, 7:17--61, 1960.

\bibitem{stoch_web}
P.~Raghavan, R.~Kumar, S.~Rajagopalan, D.~Sivakumar, A.~Tomkins, and E.~Upfal.
\newblock Stochastic models for the web graph.
\newblock {\em Proceedings of the IEEE Symposium on Foundations of Computer
  Science}, 2000.

\bibitem{strogatz_review}
S.~Strogatz.
\newblock Exploring complex networks.
\newblock {\em Nature}, 410:268--276, 2001.

\bibitem{huberman_percolation}
T.~H. T~Shraeger and B.~Huberman.
\newblock Observation of phase transitions in spreading activation networks.
\newblock {\em Science}, 236:1092, 1987.

\bibitem{fell}
A.~Wagner and D.~Fell.
\newblock The small world of metabolism.
\newblock {\em Nature Biotechnology}, 18:1121--1122, November 2000.

\bibitem{cascade}
D.~J. Watts.
\newblock A simple model of fads and cascading failures.
\newblock {\em
  http://www.santafe.edu/sfi/publications/Abstracts/00-12-062abs.html}, 2000.

\bibitem{wattssw}
D.~J. Watts and S.~Strogatz.
\newblock Collective dynamics of small world networks.
\newblock {\em Nature}, 393:440--442, 1998.

\end{thebibliography}
\bibliographystyle{abbrv}
\end{document}